\documentclass[11pt]{article}

\usepackage[margin=1in]{geometry}
\usepackage{amssymb,amsmath}
\usepackage{booktabs}
\usepackage{hyperref}
\usepackage{microtype}
\usepackage{enumitem}

\newcommand{\Pp}{\mathbb{P}}
\newcommand{\Ee}{\mathbb{E}}
\newcommand{\Dir}{\mathrm{Dir}}
\newcommand{\Mult}{\mathrm{Mult}}
\newcommand{\Beta}{\mathrm{Beta}}

\title{Exact, narrow and fast:\\ Use the Gaffke confidence interval\\ for the mean of a discrete bounded variable}
\author{Louis Abraham}
\date{}

\begin{document}
\maketitle

\begin{abstract}
Survey ratings, test scores and many other data take a few known values, and we often want a confidence interval for their mean.
For small samples, or for means close to the smallest or largest value, the usual intervals cover the true mean much less often than they claim.
The interval of Gaffke generalizes the Clopper--Pearson interval to any finite set of values, and a recent proof of Gaffke's conjecture shows that its coverage is at least the nominal level for every sample size and every distribution.
Its endpoints, however, are quantiles of a weighted sum of Dirichlet variables, with no closed form.

This note makes three contributions.
\begin{itemize}[nosep,leftmargin=1.5em]
  \item \emph{A fast and accurate algorithm for the endpoints.} A Fourier inversion integral on a line through the saddle point gives tail probabilities with a small relative error; a Gauss--Legendre rule on geometric panels converges at the same rate wherever the singularities are; a Newton iteration in the log-distance to the end of the tail needs $2.6$ evaluations per endpoint on average. The cost is $O(kN)$ operations for $k$ values and $N$ quadrature nodes, and $N$ does not grow with the sample size: an interval takes less than a millisecond for any $n$.
  \item \emph{An experimental comparison with the published intervals.} For $n \le 30$, with the coverage computed by enumeration of all outcomes, the Gaffke interval is the shortest of the intervals that keep the level: $17$ to $20\%$ shorter than the betting interval of Waudby-Smith and Ramdas, and $19$ to $37\%$ shorter than Anderson's. The shorter intervals (Wald, Student, profile likelihood, Bayesian) cover the mean less than $50\%$ of the time for some distributions.
  \item \emph{Tested Python and R packages, and a web page.} The endpoints agree with exact references (a B-spline formula in rational arithmetic, high-precision integrals) to $10^{-12}$ of the range of the values, with tail probabilities of relative error below $10^{-10}$ down to $\alpha = 10^{-12}$.
\end{itemize}
So the interval is exact, narrow and fast, and we recommend it.
\end{abstract}

\section{Introduction}

\paragraph{Problem.}
A sample of $n$ trials, each with one of $k$ known values, has an unknown mean $\mu$.
Typical examples are ratings on a scale from $1$ to $5$, scores, and counts of items in classes with known sizes.
The usual intervals for $\mu$ (Wald, Student) rest on the normal approximation, which fails for small $n$ or when $\mu$ is close to an end of the range.
Distribution-free intervals (Hoeffding, empirical Bernstein, Anderson, betting) keep the level, but they are wide.

\paragraph{What is known.}
Gaffke \cite{gaffke2005} proposed a test of the mean of a nonnegative variable and conjectured that it is valid.
Learned-Miller and Thomas \cite{lmt2019} studied the confidence interval that it gives for bounded data; for Bernoulli data it is the Clopper--Pearson interval.
Vlassis and Thomas \cite{vlassis2026} recently proved the conjecture, so the interval has guaranteed coverage, and Ming et al.\ \cite{ming2026} showed that its width is asymptotically that of the normal interval.

\paragraph{Contributions.}
\begin{enumerate}[nosep]
  \item An algorithm that computes the endpoints to $10^{-12}$ of the range of the values in $O(kN)$ operations, with a number $N$ of quadrature nodes that does not grow with $n$ (Section~\ref{sec:fast}). Its key points are the saddle-point line, which gives a small \emph{relative} error on the tail probabilities, and geometric panels, on which the Gauss--Legendre rule converges at a rate that does not depend on the data.
  \item An experimental comparison of the coverage and width of the Gaffke interval with eight published intervals, for $n \le 30$ (Section~\ref{sec:narrow}).
  \item Python and R packages, both named \texttt{gaffke}, tested against exact references and in $40{,}000$ random problems (Section~\ref{sec:tests}), and a web page that computes the interval in the browser.
\end{enumerate}
Section~\ref{sec:interval} defines the interval, and Section~\ref{sec:exact} recalls why its coverage is guaranteed.

\section{The interval}\label{sec:interval}

\paragraph{Data.}
Each of $n$ independent trials gives one of $k$ known values $v_1, \dots, v_k$, with unknown probabilities $p_1, \dots, p_k$.
The counts are $X \sim \Mult(n, p)$, and we want a confidence interval for the mean
\[
  \mu = v^\top p = \sum_i v_i p_i .
\]
We write $v_{\min}$ and $v_{\max}$ for the smallest and largest values, and $e_i$ for the $i$-th unit vector.

\paragraph{Gaffke's test.}
Let $Z_1, \dots, Z_n$ be independent nonnegative variables.
To test whether their means are at most $1$, Gaffke \cite{gaffke2005} adds a zero to the sample and averages the $n + 1$ values with random weights $D = (D_0, D_1, \dots, D_n)$ drawn from the uniform Dirichlet law $\Dir(1, \dots, 1)$.
The test rejects at level $\alpha$ when this random average is rarely at most $1$:
\[
  \text{reject when} \quad G(z) = \Pp_D\Bigl\{\sum_{i=1}^n z_i D_i \le 1\Bigr\} \le \alpha .
\]
Gaffke conjectured that this test is valid.
Vlassis and Thomas \cite{vlassis2026} recently proved it: if the $Z_i$ are independent, nonnegative and $\Ee Z_i \le 1$, then $\Pp\{G(Z) \le \alpha\} \le \alpha$ for every $\alpha$.

\paragraph{From the test to the interval.}
For a candidate mean $m > v_{\min}$, apply the test to $Z_i = (v_{I_i} - v_{\min})/(m - v_{\min})$, where $I_i$ is the category of trial $i$.
Trials in the same category have the same value, so their Dirichlet weights add up; the weight $D_0$ goes to $v_{\min}$.
So the test does not reject $m$ at level $\alpha/2$ exactly when $m$ is above the $\alpha/2$-quantile of $v^\top P_-$, where $P_-$ has the Dirichlet law with the counts as parameters and one more unit on a smallest value.
The same argument with $v_{\max} - v$ gives the upper endpoint.
The interval of level $1 - \alpha$ is
\[
  L = Q_{\alpha/2}\bigl(v^\top P_-\bigr), \quad U = Q_{1-\alpha/2}\bigl(v^\top P_+\bigr), \qquad
  P_- \sim \Dir(X + e_{i_{\min}}), \quad P_+ \sim \Dir(X + e_{i_{\max}}),
\]
where $Q_q$ is the quantile function and $i_{\min}$, $i_{\max}$ are indices of a smallest and of a largest value.

\section{Exact}\label{sec:exact}

The word \emph{exact} has here the meaning that it has for the Clopper--Pearson interval: the coverage is guaranteed for every sample size, not only in the limit.
By the result of Vlassis and Thomas, the interval covers $\mu$ with probability at least $1 - \alpha$, for every $p$ and every $n$.
Before this proof, Learned-Miller and Thomas \cite{lmt2019} had proved it for Bernoulli and half-Bernoulli data.
The interval generalizes the Clopper--Pearson interval \cite{clopper1934}: for the values $v = (0, 1)$ and $x$ successes in $n$ trials, $v^\top P_- \sim \Beta(x, n - x + 1)$ and $v^\top P_+ \sim \Beta(x + 1, n - x)$, so
\[
  L = I^{-1}_{\alpha/2}(x,\, n - x + 1), \qquad U = I^{-1}_{1 - \alpha/2}(x + 1,\, n - x),
\]
which are the Clopper--Pearson bounds ($I^{-1}$ is the inverse of the regularized incomplete Beta function).
In our experiments (Section~\ref{sec:tests}), some one-sided coverages are equal to $1 - \alpha/2$, so the bound cannot be improved in general.

\section{Narrow}\label{sec:narrow}

Any interval can be made valid by making it wide enough; the question is how wide a valid interval must be.
Table~\ref{tab:coverage} gives the coverage of published intervals for three values $v = (0, \frac12, 1)$ and $\alpha = 0.05$: the minimum over a lattice of $3321$ points $p$ of the simplex, boundary included, and the mean over the lattice of the expected width.
The coverage is computed by enumeration of all outcomes.
The intervals above the line keep the level (the value $0.946$ of the betting interval comes from the approximations given in the caption): Gaffke's is the shortest of them, by $17$ to $20\%$ against the betting interval of Waudby-Smith and Ramdas \cite{wsr2024} and by $19$ to $37\%$ against Anderson's \cite{anderson1969}.
The intervals below the line are shorter by $5$ to $30\%$, but their coverage falls far below $95\%$ for some $p$.
The results are the same for the values $(0, 0.1, 1)$ and for five values.

For large $n$ the interval is as narrow as the normal interval: $\sqrt n$ times its width tends to $2\sigma z_{1-\alpha/2}$ almost surely \cite{ming2026}.
It is not the shortest valid interval for every outcome: Ming et al.\ construct a valid interval that is strictly shorter in some cases, but no published interval is shorter in our comparison.

\begin{table}[h]
\centering
\small
\begin{tabular}{lrrrrrr}
\toprule
& \multicolumn{3}{c}{minimum coverage} & \multicolumn{3}{c}{mean width} \\
\cmidrule(lr){2-4}\cmidrule(lr){5-7}
Interval & $n = 5$ & $10$ & $30$ & $5$ & $10$ & $30$ \\
\midrule
Gaffke & 0.976 & 0.963 & 0.954 & 0.648 & 0.472 & 0.268 \\
Betting$^{*}$ & 0.991 & 0.982 & 0.946 & 0.797 & 0.592 & 0.323 \\
Anderson & 0.991 & 0.992 & 0.991 & 0.802 & 0.648 & 0.426 \\
Hoeffding & 0.991 & 0.992 & 0.991 & 0.878 & 0.728 & 0.470 \\
Empirical Bernstein & 1.000 & 1.000 & 1.000 & 1.000 & 1.000 & 0.847 \\
\midrule
Student $t$ & 0.061 & 0.118 & 0.314 & 0.619 & 0.451 & 0.251 \\
Wald & 0.061 & 0.118 & 0.314 & 0.454 & 0.378 & 0.237 \\
Profile likelihood & 0.446 & 0.477 & 0.493 & 0.474 & 0.377 & 0.236 \\
Bayes, Jeffreys prior & 0.000 & 0.000 & 0.000 & 0.490 & 0.377 & 0.234 \\
\bottomrule
\end{tabular}

\caption{Coverage and mean expected width, $v = (0, \frac12, 1)$, $\alpha = 0.05$ (nominal coverage $0.95$). Wald: mean $\pm z\,\hat\sigma/\sqrt n$; Hoeffding \cite{hoeffding1963}; empirical Bernstein \cite{maurer2009}; Anderson: the one-sided DKW band on the empirical distribution function \cite{anderson1969}; profile likelihood with the $\chi^2_1$ cutoff; Bayes: equal-tail credible interval with the prior $\Dir(\frac12, \dots, \frac12)$. $^{*}$The betting interval depends on the order of the trials; its coverage is averaged over $8$ random orders, and computed on a grid of step $10^{-3}$.}
\label{tab:coverage}
\end{table}

\section{Fast}\label{sec:fast}

The endpoints have no closed form, and a level such as $1 - 10^{-6}$ needs tail probabilities of $5 \times 10^{-7}$ with a small relative error.
This section gives the method (Sections~\ref{sec:red}--\ref{sec:quantile}) and its cost (Section~\ref{sec:cost}).
Each endpoint is a quantile of
\[
  Y = w^\top P, \qquad P \sim \Dir(a), \qquad (w, a) = (v,\ X + e_{i_{\min}}) \text{ or } (v,\ X + e_{i_{\max}}).
\]

\subsection{Reductions}\label{sec:red}

Drop the categories with $a_i = 0$, merge equal values, and scale the values to $[0, 1]$:
\[
  a'_j = \sum_{i : w_i = u_j} a_i, \qquad w'_j = \frac{u_j - w_{\min}}{w_{\max} - w_{\min}},
\]
where $u_1 < \dots < u_{k'}$ are the distinct values (sums of Dirichlet coordinates are Dirichlet).
All tolerances are relative to $w_{\max} - w_{\min}$.
For $k' = 1$, $Y = w_{\min}$; for $k' = 2$,
\[
  Y = w_{\min} + (w_{\max} - w_{\min})\, B, \quad B \sim \Beta(a'_2, a'_1), \qquad
  Q_q(Y) = w_{\min} + (w_{\max} - w_{\min})\, I^{-1}_q(a'_2, a'_1),
\]
as for the Clopper--Pearson bounds.
Below, $k' \ge 3$ and we drop the primes.

\subsection{Tail probability as an integral}

With independent $\Gamma_i \sim \mathrm{Gamma}(a_i, 1)$, $P = \Gamma/\sum_i \Gamma_i$, so for $0 < y < 1$
\[
  \Pp(Y \le y) = \Pp(S \le 0), \qquad S = \sum_i \lambda_i \Gamma_i, \quad \lambda_i = w_i - y,
\]
\[
  M(s) = \Ee\, e^{sS} = \prod_i (1 - s\lambda_i)^{-a_i}, \qquad \kappa(s) = \log M(s), \qquad \frac{1}{\lambda_{\min}} < s < \frac{1}{\lambda_{\max}} .
\]
For a real $c \neq 0$ in this interval, the inversion formula on the line $\operatorname{Re} s = c$ gives \cite{gilpelaez1951}
\begin{equation}\label{eq:inv}
  \Pp(S > 0) = \frac1\pi \int_0^\infty \operatorname{Re}\frac{M(c+it)}{c+it}\,dt \ \ (c > 0), \qquad
  \Pp(S \le 0) = -\frac1\pi \int_0^\infty \operatorname{Re}\frac{M(c+it)}{c+it}\,dt \ \ (c < 0).
\end{equation}
We compute the smaller tail: $\Pp(S \le 0)$ if $\Ee S = \sum_i a_i\lambda_i > 0$, else $\Pp(S > 0)$.
We take $c$ at the saddle point \cite{daniels1954,helstrom1978},
\[
  \kappa'(c) = \sum_i \frac{a_i \lambda_i}{1 - c\lambda_i} = 0,
\]
found by Newton's method with bisection, stopped when $|\kappa'(c)| \le 0.1\sqrt{\kappa''(c)}$, and moved away from $0$ if $|c| < \min(1, |c_{\mathrm{pole}}|/2)$ (scaled as below).
With $p_i = 1 - c\lambda_i > 0$ and $x_i(t) = t\lambda_i/p_i$,
\[
  M(c + it) = e^{\kappa(c)}\, r(t)\, e^{i\theta(t)}, \qquad
  r(t) = \prod_i (1 + x_i^2)^{-a_i/2}, \qquad \theta(t) = \sum_i a_i \arctan x_i ,
\]
so the two integrals of~\eqref{eq:inv} have the integrand
\begin{equation}\label{eq:tail}
  \frac{e^{\kappa(c)}}{\pi}\, r(t)\, \frac{c\cos\theta(t) + t\sin\theta(t)}{c^2 + t^2} .
\end{equation}
Its size at $t = 0$ is $e^{\kappa(c)}/(\pi|c|)$, the size of the tail, so a quadrature error small relative to the integral is small relative to the tail.
The derivative in $y$ gives the density with the same $r$ and $\theta$:
\begin{equation}\label{eq:pdf}
  f_Y(y) = \frac{e^{\kappa(c)}}{\pi} \int_0^\infty r(t) \Bigl(\cos\theta(t) \sum_i \frac{a_i/p_i}{1 + x_i^2} - \sin\theta(t) \sum_i \frac{a_i x_i/p_i}{1 + x_i^2}\Bigr)\,dt .
\end{equation}
The code returns $\kappa(c) + \log(\cdot)$ of both integrals.
All of this is computed in the variable $s/\sigma$, with $\sigma^2 = \sum_i a_i\lambda_i^2$, so that the integrand changes on a scale of about $1$.

\subsection{Quadrature}

The singularities of the integrands, in the variable $t$, are the branch points and the pole,
\[
  t = -i\,\frac{p_j}{\lambda_j} \ \ (j = 1, \dots, k), \qquad t = i c,
\]
all on the imaginary axis.
We cut the integral at $T$ and split $[0, T]$ into panels whose lengths double:
\[
  \tfrac12\sum_i a_i \log\bigl(1 + T^2\lambda_i^2/p_i^2\bigr) = 40, \qquad
  [0, T] = [0, h] \cup \bigcup_{j \ge 0} [2^j h, 2^{j+1} h], \qquad
  h = \min\Bigl(1, \tfrac12 |c|, \tfrac12 \min_j \tfrac{p_j}{|\lambda_j|}\Bigr),
\]
so that $r(T) = e^{-40}$, with the last panel ending at $T$.
On each panel $[x_0, x_1]$ we use the 12-point Gauss--Legendre rule,
\[
  \int_{x_0}^{x_1} \phi(t)\,dt \approx \frac{x_1 - x_0}{2} \sum_{l=1}^{12} \omega_l\, \phi\Bigl(\frac{x_0 + x_1}{2} + \frac{x_1 - x_0}{2}\xi_l\Bigr).
\]
If $\phi$ is analytic and bounded by $B$ inside the Bernstein ellipse $E_\rho$ of the panel, the error is at most \cite[Thm.~19.3]{trefethen2013}
\[
  \frac{64}{15}\, \frac{B\, \rho^{-24}}{\rho^2 - 1}\cdot\frac{x_1 - x_0}{2} .
\]
For $[x, 2x]$, the imaginary axis is the line $\operatorname{Re} u = -3$ in the normalized variable $u \in [-1, 1]$, so every $\rho < 3 + 2\sqrt 2$ is allowed, for every panel and every position of the singularities: the rule converges at the same rate on all panels.
When the parameters have very different sizes, $\theta$ can turn fast far from $t = 0$; we split a panel of length $\ell$ into $\lceil \ell \max|\theta'| / 6 \rceil$ equal parts, with
\[
  \theta'(t) = \sum_i \frac{a_i \lambda_i/p_i}{1 + x_i^2} ,
\]
so that $\theta$ turns by at most $3$ radians on half of each part.

\subsection{Quantile}\label{sec:quantile}

To solve $\mathrm{tail}(y) = q$, let $d$ be the distance from $y$ to the end of the tail ($d = y$ or $d = 1 - y$) and $u = \log d$.
Near $d = 0$, $\mathrm{tail} \approx C d^{\,a}$, so $\log \mathrm{tail}$ is nearly linear in $u$, and we use Newton's method on
\[
  g(u) = \log \mathrm{tail}(y) - \log q, \qquad g'(u) = \frac{d\, f_Y(y)}{\mathrm{tail}(y)}, \qquad u \leftarrow u - \frac{g(u)}{g'(u)},
\]
with $f_Y$ from~\eqref{eq:pdf}, inside a bracket $[d_{\mathrm{lo}}, d_{\mathrm{hi}}]$ of the root (bisection when a step leaves it).
The start point is the quantile of the Beta law with the mean and variance of $Y$:
\[
  m = \frac{\sum_i a_i w_i}{a_0}, \quad
  \mathrm{Var}\,Y = \frac{\sum_i a_i (w_i - m)^2}{a_0(a_0 + 1)}, \quad
  \nu = \frac{m(1-m)}{\mathrm{Var}\,Y} - 1, \quad
  y_0 = I^{-1}_q\bigl(m\nu, (1 - m)\nu\bigr),
\]
with $a_0 = \sum_i a_i$; it is exact for two values.
The iteration stops when $|\Delta d| \le \epsilon$ or when the predicted error of the next step is small,
\[
  100\, d \,\Bigl|\frac{g''}{2g'}\Bigr| (\Delta u)^2 \le \epsilon, \qquad g'' \approx \frac{g'(u) - g'(u_{\mathrm{prev}})}{u - u_{\mathrm{prev}}},
\]
with $\epsilon = 10^{-12}$ by default.
The upper endpoint solves $\Pp(Y > y) = \alpha/2$, so $1 - \alpha/2$ is never formed.
\subsection{Cost}\label{sec:cost}

Let $E$ be the number of evaluations of the tail and density, $I$ the number of Newton steps of one saddle point search, and $N$ the number of quadrature nodes of one evaluation.
With $J = \lceil \log_2(T/h) \rceil$ panels before the phase splits,
\[
  \text{cost of one endpoint} = O\bigl(k\log k + E\,(I + N)\,k\bigr), \qquad N = 12\,(J + 1) + 12 \times \text{(splits)} .
\]
The sample size $n$ enters only through the parameters $a$, and the cost does not grow with it.
When $a_0$ is large, $Y$ is close to normal, the integrand is close to $e^{-t^2/2}$ in the scaled variable, and $T$, $h$ and $N$ tend to constants: $N = 60$ (five panels) for $n \ge 10^6$ in our tests.
Small parameters give a slower decay of the integrand, so a larger $T$, but only $\log_2 T$ more panels.
We have no proven bound on $N$; over random problems with $k \le 20$, $n$ from $5$ to $10^9$ and $\alpha \in \{0.05, 10^{-6}\}$, the median of $N$ is $60$ to $192$ and the maximum is $456$.
In our tests $E$ is $2.6$ on average and at most $7$, and $I$ is a few steps, since each search starts from the last saddle point.

For comparison, a Monte Carlo estimate of a tail probability $q$ with relative error $\delta$ needs about $1/(q\delta^2)$ samples, which is $2 \times 10^{26}$ for $q = 5 \times 10^{-7}$ and $\delta = 10^{-10}$.
An interval takes $0.15$ to $0.3$~ms for $k \ge 3$ values and about $0.02$~ms for two, almost all of it Python overhead.

\section{Accuracy}\label{sec:tests}

We have no proof of an error bound: the argument on the panels gives the rate of convergence, and the tests measure the error.
We compare the code with these references:
\begin{itemize}[nosep]
  \item \emph{Exact law for integer parameters.} $Y$ has the law of a B-spline with knots $w_i$ of multiplicities $a_i$ \cite{curry1966}, and with $m = a_0 - 1$,
  \[
    \Pp(Y > y) = \sum_{j : w_j > y} \operatorname{Res}_{t = w_j} \frac{(t - y)^m}{\prod_l (t - w_l)^{a_l}},
  \]
  which we compute in exact rational arithmetic.
  \item \emph{High precision.} The Beta law for two values, and the inversion integral with 40 digits for large or very unequal parameters.
  \item \emph{Simulation and properties.} Monte Carlo samples; invariance under affine maps of the values, permutations and splits of a category; monotonicity in $\alpha$; edge cases such as $n \le 1$, $n = 10^{12}$ and $\alpha = 10^{-15}$.
  \item \emph{Coverage.} The coverage of the interval, computed by enumeration of all outcomes, for three values and $n \le 13$.
\end{itemize}

In all tests, including $40{,}000$ random problems with up to $20$ values and $n$ up to $3 \times 10^6$:
\begin{itemize}[nosep]
  \item the endpoints are within $10^{-12}$ of the range of the values;
  \item the tail probabilities have a relative error below $10^{-10}$, down to $\alpha = 10^{-12}$;
  \item each one-sided coverage is at least $1 - \alpha/2$.
\end{itemize}

\section*{Software}

\begin{itemize}[nosep]
  \item Python: \texttt{pip install gaffke}, \url{https://pypi.org/project/gaffke/}
  \item R: \texttt{install.packages("gaffke")}, \url{https://CRAN.R-project.org/package=gaffke}
  \item Web: \url{https://louisabraham.github.io/gaffke/}, which computes the interval in the browser with a JavaScript port (\texttt{gaffke.js}) that agrees with the Python package to $10^{-15}$ of the range of the values
\end{itemize}
Both give \texttt{gaffke\_ci(counts, values, alpha)} and the law of $w^\top P$ (distribution function, density, quantile).
The R package is a line-by-line port of the Python one and agrees with it to $10^{-10}$ of the range of the values.
The source code, the tests, the experiments of Section~\ref{sec:narrow} and this note are at \url{https://github.com/louisabraham/gaffke}, under the MIT license.

\end{document}